# Thoughts on opportunities in high-energy nuclear collisions

*Federico Antinori, Peter Braun-Munzinger, Jan-Fiete Grosse-Oetringhaus, Ulrich Heinz, Barbara Jacak, Peter Jacobs, Alexander Kalweit, Volker Koch, Yen-Jie Lee, Marco Van Leeuwen, Silvia Masciocchi, Guilherme Teixeira De Almeida Milhano, Alexander Milov, Andreas Morsch, Berndt Mueller, James Lawrence Nagle, Antonio Ortiz, Guy Paic, Dennis Perepelitsa, Krishna Rajagopal, Ralf Rapp, Gunther Roland, Paul Romatschke, Jurgen Schukraft, Yves Schutz, Johanna Stachel, Xin-Nian Wang, Urs Achim Wiedemann, Zhangbu Xu, William Zajc.*

## Abstract

*This document reflects thoughts on opportunities from high-energy nuclear collisions in the 2020's.*

## Introduction

On the week-end of May 19-21, 2018, a group of 30 physicists met in the monastery at Terzolas in the Italian Alps to discuss recent developments in the field of ultra-relativistic heavy-ion physics and the future opportunities arising from them. These discussions took place, in the Mont Sainte Odile meetings series, for the fourth time on the weekend after a Quark Matter conference. The Terzolas meeting focused on four subject areas that had featured prominently in recent scientific debates:

1. <u>Collectivity in small systems</u>: Does the discovery of signatures of collectivity in small (proton-nucleus and proton-proton) collision systems call into question the prevailing fluid dynamic interpretation of soft particle production in nucleus-nucleus collisions? How can these observations be employed further to understand the dynamical origin of the observed collectivity?
2. <u>Physics interpretation of hard probes</u>: Given the recent progress in modern jet and jet substructure measurements in nucleus-nucleus collisions, how can these techniques help to characterize medium properties and give access to the microscopic structure of the produced matter? What is the role of heavy flavor in this program of probing the produced dense QCD matter with hard processes?
3. <u>Electromagnetic probes</u>: From observing the electromagnetic radiation of a thermalized quark gluon plasma to characterizing signatures of chiral symmetry restoration in dilepton spectra, the measurement of electromagnetic processes is clearly motivated by fundamental questions in nucleus-nucleus collisions. What are the perspectives for experimental and theoretical advances in the coming years?
4. <u>Soft precision physics</u>: Despite the increasingly complete characterization of soft particle production in all collision systems and all experimentally accessible center-of-mass energies, there are motivations for exploring this physics at higher precision and with refined observables that range from characterizing the QCD phase diagram with fluctuation measurements to open questions in hadrochemistry to understanding the production mechanisms of nuclei and anti-nuclei in kinematic ranges relevant for ultra-energetic cosmic rays. What is needed to make further progress in these directions?

As in previous meetings, all participants met as individual scientists without any mandate from a collaboration or from their scientific community, and without any firmly defined agenda of what must and what should not be covered in two days of intense debate. The participants were asked prior to the meeting to formulate their considerations in the form of itemized statements. These inputs were grouped by topic and they were summarized during the meeting by conveners in focused 10-minutes interventions that were followed by free-ranging exchanges amongst all participants. While this short report of the main lines of discussion will hardly capture the non-conventional character of the debate, it serves a dual purpose: First, for the participants, it is a monastic effort of establishing to what extent their wide-ranging discussion has reached conclusions. Second, this write-up aims, of course, at contributing to the current debate of how to best exploit the scientific opportunities arising from HL-LHC and the future RHIC running.

## Collectivity in Small Systems

The growing body of evidence that small collision systems appear to exhibit flow signals of similar character and magnitude as those found in large systems was much discussed. The fluid paradigm via which a by now enormous and varied suite of experimental observables in heavy-ion collisions are described in terms of the formation of a droplet of liquid quark-gluon plasma that expands and cools according to the laws of relativistic dissipative hydrodynamics and subsequently falls apart into an explosion of hadrons has been described and assessed in detail in a previous Mont Sainte Odile meeting. The write-up [1] of this previous meeting includes a description of the kinds of observations that support the paradigm, and a discussion of further experimental tests. Already at that time, the occurrence of flow-like phenomena in small — but dense — systems probed in high-multiplicity proton-nucleus and proton-proton collisions was regarded as surprising, and it was noted that it raised fundamental questions about how small a droplet can be, and how dense a droplet must be, in order for the fluid paradigm to be applicable. The discussion in Terzolas also focused on the challenges to the application of the fluid paradigm in small dense systems posed by the absence of possible accompanying signatures of parton energy loss, as commonly understood from the phenomenology of large systems. In data from small systems, all effects thought to be related to jet quenching are to date too small to be seen within present experimental sensitivity. Furthermore, experimental evidence of flow signals has not (yet) been observed in even smaller systems, such as those accessible in archived electron-proton and $e^+e^-$ collision data, and it remains to be clarified how to quantify whether flow signals are present in the most peripheral heavy-ion collisions. These findings, in sum, raise deep questions about how features of the data traditionally understood as arising from flow, and parton energy loss, are to be interpreted in terms of their underlying physical origin, and the system sizes or regimes over which they are present.

To address this, studies of existing small-systems data should be intensified to increase the significance of the experimental conclusions, and new techniques should be devised which are aimed at either establishing energy loss signatures and then characterizing them, or at setting robust limits on their existence. Concerning experimental progress on collectivity in small systems, the discussion emphasized the much-improved characterization of flow-like phenomena in the small proton-proton and proton-nucleus collision systems achieved over the last two years. It was noted, however, that this is still far from the state of the art reached in nucleus-nucleus collisions. In particular, information about correlations between different flow coefficients and reaction plane correlations are still scarce in proton-nucleus and absent in proton-proton collisions, and such information would provide valuable dynamical tests as it gives access to non-linear response coefficients. It will be interesting to see whether these correlations can be isolated and distinguished from other correlations and fluctuations in proton-proton and/or proton-nucleus collisions.

Recent developments in the foundations of fluid dynamics, particularly in the discovery of hydrodynamic attractors, puts the applicability of hydrodynamics to, and breakdown in, small systems on a firm theoretical footing. There has been much progress in recent years in understanding how initially far-from-equilibrium, anisotropic, expanding systems hydrodynamize. An understanding of the system-size dependence of these mechanisms seems within reach. That said, additional theoretical and experimental efforts will be needed to quantitatively constrain the hydrodynamic breakdown scale in QCD, and thus the characteristics of the smallest possible droplet of QCD fluid.

On the phenomenological level, fluid dynamic simulation codes have reached a certain level of maturity in recent years. The current state of the art in heavy-ion collisions involves comparison to a broad suite of experimental observables constituting a global description of soft particle production. And, the comparison to data is now done with Bayesian techniques that quantify uncertainties in the conclusions drawn, allowing assessments of uncertainties in the determination of the initial conditions and the transport properties of the hydrodynamic fluid, both, via common analyses of diverse data sets. It is an open challenge to extend studies at this level of sophistication to the smaller systems, allowing precise measurement and global analyses of spectra, momentum-dependent flow, and – when they have been measured – correlations to test the system-size dependence of conclusions drawn regarding the properties of the QGP fluid, and regarding the applicability of the fluid paradigm. In tandem, reliable theoretical calculations of the expected magnitude of energy loss effects in small systems, and novel signatures through which they may be observed, are needed to guide experimental efforts on this front. Particular emphasis should be placed on how the body of experimental data and theoretical guidance are to be understood alongside other signatures such as quarkonia suppression or regeneration, thermal radiation, the flow of heavy quarks, and azimuthal anisotropies at very large transverse momentum.

Non-hydrodynamic explanations for experimental flow signatures are being explored and offer intriguing and different pictures of the underlying microscopic dynamics. Two classes of such proposed explanations that have been discussed in detail are transport-based approaches and approaches based on initial state correlations based on glasma diagrams. It was mentioned at the meeting that even in the absence of a saturated initial parton density, non-vanishing contributions to higher-order cumulant flow signals may arise due to interference-effects in multi-parton final states. The good phenomenological agreement of some glasma models with some data was noted. However, it remains to be better understood to what extent agreement between these models and data is unambiguous, as would be possible in a model based only upon controlled approximations and what role uncertainties in the phenomenological modeling play. The discussion at Terzolas focused more on transport-based approaches.

One entry to the discussion of transport-based approaches, in which partons produced in the collision fly for some distance between discrete scattering events, begins by noting the marked qualitative differences between the fluid paradigm (in which soft particle production in heavy-ion collisions is preceded by a hydrodynamic epoch) and the models implemented in the standard general-purpose event generators used to simulate soft particle production in proton-proton collisions. The latter models essentially assume that particles, once produced, free-stream to the detector. They may fragment while free-streaming, but they do not undergo further interactions with other particles produced initially, and no flow is generated. From this starting point, adding final-state interactions collision-by-collision, as in a transport model, seems like a natural next step in the discussion of small collision systems in which the final state is dense. Free-streaming (as in conventional approaches to proton-proton collisions) means that the mean free path $l_{mfp}$ of the partons produced in the collision is assumed to be significantly larger than the system size R. In contrast, a hydrodynamic approach becomes valid whenever $l_{mfp}$ is significantly smaller than R, and the dissipation during the hydrodynamic flow becomes small (as in the fluid dynamics of heavy-ion collisions) when $l_{mfp}$ is not much larger than the microscopic length scale set by the energy density of the fluid, namely 1/T at equilibrium. Note that holographic fluids, in which the coupling is strong enough that there are no well-defined quasiparticles at all and so in reality no mean free path can be defined can, somewhat loosely, be thought of as fluids in which $l_{mfp}$ is comparable to $1/\pi T$. If a system is dense enough, or large enough, that 1/T (equivalently, the inverse energy density to the one fourth power) is much smaller than R, then the question of whether a hydrodynamic approach is valid depends on $l_{mfp}$. At lower densities, hydrodynamics cannot be applied. This discussion highlights the fact that in order to understand how collectivity in small (proton-proton, proton-nucleus) collision systems arises, and how its analysis can help to constrain our dynamical understanding of collectivity in nucleus-nucleus collisions, we must understand whether these systems differ mainly in their transverse extent or whether there are also significant differences in their intrinsic properties, in particular their local energy density, but ideally also their $l_{mfp}$. It is an open challenge to find observables that allow comparisons to be made between proton-proton, proton-nucleus, and nucleus-nucleus collisions with the same energy density but differing transverse size, or with the same transverse size but differing energy density.

Transport models have predicted or reproduced some, or even many, experimental flow measurements, although none provide a global description of soft particle production. Interpretation of their phenomenological successes must take into account the fact that transport models in general require additional external input or assumptions concerning the microscopic degrees of freedom, and therefore include different physics effects that go beyond parton-parton collisions. For example, the AMPT model includes concepts like *string melting* to define the initial parton distribution. It remains to be better understood to what extent agreement between these models and data is unambiguous, and what role uncertainties in the phenomenological modeling play. From the study of simpler transport models that invoke only collisions, to dedicated runs of more complex codes in which certain assumptions are switched off or altered, to studies that analyze as a function of system size how transport models hydrodynamize, to more complete model comparisons with more classes of data in particular including unified descriptions of flow and parton energy loss in transport-based approaches, there is a range of questions on which progress in our theoretical understanding should be possible.

Since transport theory can interpolate between free streaming and viscous fluid dynamics in this way, it provides an appropriate setting for understanding the dependence of fluid-like behavior on system size in proton-proton, proton-nucleus and very peripheral nucleus-nucleus collisions even at collision energies where $l_{mfp}$ is comparable to R and fluid dynamic evolution may not be taken for granted. It will therefore be interesting to further investigate whether there are generic features of kinetic transport that can account for qualitative properties of the system size dependence in these collisions.

Future research remains to establish to what extent transport theory can provide a phenomenologically valid description of soft particle production in collision systems that are small enough and have a low enough density that $l_{mfp}$ is comparable to R and that have $l_{mfp} \gg 1/T$, meaning that well-defined quasiparticles scatter zero or one or few times, and yet in which some flow observables are nonzero. In addition, studies of the system size dependence of transport models need to establish how, as a function of increasing system size, transport models approach hydrodynamic behavior, and whether transport descriptions applied to such sufficiently large systems are consistent with the properties of liquid QGP inferred from viscous hydrodynamic analyses that can be extended into regimes in which $l_{mfp}$ becomes comparable to 1/T and quasiparticles can no longer be found. As transport theory formulates a combined dynamical framework for fluid dynamic and particle-like excitations, progress in these directions should shed light on how the latter start dominating over the former as a function of decreasing system size.

# Hard Probes

The focus of the characterization of quenching phenomena in hard processes has evolved in recent years. Initial studies of jet quenching focused on inclusive production and correlations of high-$p_T$ hadrons, while the techniques to control the large backgrounds underlying reconstructed jet measurements in such complex collisions took over a decade to master. In recent years novel tools and observables to characterize the jet substructure and the correlations between the jet and the medium in which it is embedded have gained extensive interest, driven by the goal to understand the energy loss mechanism and resulting modifications of the parton shower in detail. This knowledge can subsequently be leveraged to study the short-length scale structure of the QGP. This evolving approach is accompanied by developments of a broad set of simulation tools that are new for the field of ultra-relativistic heavy-ion collisions and that – as far as they are adopted from high-energy physics – need to be rethought and appropriately modified. This includes in particular techniques for background subtraction, as well as a large suite of so-called grooming techniques.

In this context, the discussion has focused in particular on three lines of developments that are now coming in reach of joint experimental and theoretical investigation:

1. QGP response

A salient feature of jets reconstructed in heavy-ion collisions is the presence of an excess (relative to the proton-proton case) of low transverse momentum particles, inside the jet cone but also extending to large rapidity-azimuth distances from the jet axis. In very generic terms, such a particle distribution can arise from both QGP-induced modifications of the branching process and response of the QGP to the passage of, and interaction with, the jet.

While there is a broad consensus that QGP response is an intrinsic element of jet-QGP interaction, and thus an unavoidable component of reconstructed jets, its current implementations vary substantially from model to model. Compounded with model-to-model differences in the implementation of in-QGP branching dynamics, this makes the unequivocal assignment of specific jet components to QGP response difficult at best. At the experimental level, one is faced with the challenge to define and measure observables that are able to constrain the QGP-response component of jet-QGP interaction models.

The identification and study of QGP-response contributions to jets offer a unique window into the dynamics of thermalization, justifying a concerted effort to find observables most (in)sensitive to QGP response contributions. Meaningful model calculations require the inclusion of both realistic and well-justified modified branching dynamics, and realistic QGP response. Conclusions drawn from agreement of incomplete modelling (e.g. that does not account for QGP response) with experimental data are valid only if insensitivity to missing ingredients can be demonstrated. Possible paths for progress in isolating QGP response include the investigation of the energy flow in events where a jet recoils against a trigger (jet, hadron, photon, Z), the exploration of multi-particle correlations in η-φ; the exploration of the dependence of observed patterns on collision geometry, and related correlations with the event plane. Further insight may be gained through correlation studies including open heavy-flavor hadrons, once larger data samples at LHC and RHIC become available. A multi-model comprehensive scan establishing the (in)sensitivity of observables to QGP response appears essential for progress. In carrying out such a study one should also aim to identify observables whose sensitivity to background subtraction procedure is small and/or well-controlled.

## 2. Parton showers and sub-structure

Jet substructure measurements and related theoretical calculations have become a central topic of discussion in heavy-ion physics. However, a definitive case for the potential of substructure techniques to advance the understanding of QGP structure and dynamics remains to be clearly formulated. To build a clear-cut case requires the understanding of possible limitations in existing substructure methods – to a large extent imported from the high energy physics context – imposed by the presence of the large underlying event and its fluctuations. Methods to lift those limitations with heavy-ion specific tools have to be developed.

Carefully engineered tools arguably offer the most promising path to disentangle contributions to a jet originating from different mechanisms (QGP response, QGP-induced color decoherence, etc.) or to relate modifications of the jet structure at specific momentum scales to interactions with quasiparticles describing QGP excitations at corresponding length scales. A recent example in this context is given by the insight that the *Lund plane* is a measurable construct from which a variety of observables can be projected out. Combined with grooming techniques, the *Lund plane* allows for the isolation of specific emission regions which can, in turn, be related to specific jet modification mechanisms within a given model. In the short term, these recent developments should lead to the identification of observables that are largely insensitive to background fluctuations, and of others that are largely insensitive to QGP-induced jet modifications. These would constitute important control variables. Similarly, it is essential to develop observables that are sensitive to specific processes in the QGP (e.g. gluon radiation, large angle radiation, collisional broadening, etc.).

## 3. Quarkonia and open heavy flavor

The general belief that the observed relative enhancement (comparing LHC vs RHIC, low transverse momenta vs high transverse momenta, mid-rapidity vs forward rapidity systematics) of the $J/\psi$ production in nucleus-nucleus collisions signals deconfinement. The remaining challenges for this picture were discussed and the possibility of hadronic recombination (DD $\rightarrow J/\psi \, \pi$) was identified. The importance of such hadronic rescattering processes will be investigated during LHC Run 3 through precision measurements of $J/\psi$ and $\psi$' distributions, and possibly of $J/\psi - \pi$ correlations. Note that the $\psi' - \pi$ cross section is much larger than the $J/\psi - \pi$ cross section because of the very different rms radii of $J/\psi$ and $\psi$'. By detailed balance this implies also larger effects on possible $\psi$' production in the hadronic phase. Clearly, the $\psi$' over $J/\psi$ ratio versus centrality will be a key observable.

At high transverse momenta, $J/\psi$ suppression in nucleus-nucleus collision systems is usually explained by the dominance of the color screening effect. However, several observations might challenge this explanation. Neither transport models nor color screening can explain the finite elliptic flow of the $J/\psi$ at high transverse momenta. The $J/\psi$ nuclear modification factor at high transverse momenta is very similar to those of other heavy-flavor mesons and inclusive hadrons. Moreover, in proton-proton collisions the distribution of the fraction of jet transverse momentum carried by directly produced $J/\psi$ is surprisingly low and not explained by models, showing that we do not fully understand $J/\psi$ formation in pp. These observations raise the question of the role of partonic energy loss in $J/\psi$ suppression at high transverse momenta. In particular it is important to clarify the role of gluon splitting as a production mechanism for open charm and charmonia and its possible consequences for energy loss.

## Electromagnetic Probes

Electromagnetic (EM) radiation from heavy-ion collisions provides a wide range of insights into the properties of the medium produced in these reactions, including basic kinematic information (fireball temperature, degree of collectivity and lifetime) as well as dynamical information (in-medium spectral function encoding changes in degrees of freedom and the restoration of the spontaneously broken chiral symmetry).

The NA60 di-muon spectra measured in 17.3 GeV center-of-mass energy In-In collisions at the SPS are the current gold standard in the field. Due to high statistics, excellent mass resolution and displaced vertex rejection, it was possible to subtract the *cocktail* of contributions from final-state decays (including those from correlated heavy-flavor decays), fully correct for experimental acceptance and isolate the radiation spectrum from the hot and dense fireball, for invariant masses from threshold to the $J/\psi$ mass. This, in turn, provided unprecedented insights into the properties of the medium: In the low-mass region (M < 1 GeV), the melting of the $\varrho$-meson line shape was established, indicating a transition from hadronic degrees of freedom into a structure less quark-antiquark continuum and encoding the manifestation of chiral symmetry restoration. The total low-mass yield gives a direct measure of the total fireball lifetime; and the slope of the intermediate mass spectrum (1 GeV < M < 3 GeV) provides a pristine measurement of the early temperatures of the fireball (T ~ 205 MeV in the NA60 spectra), free of blue-shift effects from the collective medium expansion that affect measurements of the transverse momentum spectrum. Ideally, data on the mass spectra could be augmented by mass-binned elliptic flow coefficient measurements, to further narrow down the emission history of the radiation.

There was broad agreement that precision dilepton measurements with heavy colliding systems are a high-priority goal at RHIC and the LHC. Secondary vertex rejection will enable unique temperature measurements of the QGP, and the low-mass line shape will provide a critical test of the $\varrho$-melting scenario at (near) vanishing baryon chemical potential. This scenario has been shown to be consistent with expectations of chiral symmetry restoration via both chiral sum rules in hadronic matter and a degeneracy of hadronic and QGP rates at temperatures near 200 MeV. A precision measurement at collider energies will provide additional constraints that can be directly tested against lattice QCD (LQCD) results, to delineate the melting from other explanations such as collision broadening. Improved LQCD calculations of the vector correlator are anticipated. Measurements of simultaneous changes in the axial-vector $a_1$ meson, the chiral partner of the $\varrho$, would provide compelling evidence of chiral symmetry restoration, but are extraordinarily challenging in the nucleus-nucleus environment. It is not even clear whether it is possible to detect the $a_1$ in proton-proton collisions at collider energies, but this possibility warrants investigation, for example in the $\pi^{\pm}$ channel.

Concerning direct-photon emission, the existing tension between the low momentum spectra from PHENIX and STAR in 200 GeV Au-Au collisions, on the order of a factor of two, remains to be resolved. The PHENIX measurements also exceed those from theoretical calculations by a similar factor. A complementary STAR measurement of photon (or dilepton) elliptic flow coefficient would be very valuable. On the theoretical side, the QGP emission rates warrant further investigations of non-perturbative effects that are not included in NLO calculations (for example due to remnants of the confining interaction). Such interactions are inevitable as the QGP approaches the transition regime, and may play a role at significantly higher temperatures. In the confined phase, the effective hadronic interactions are well constrained through symmetry principles and empirical knowledge of vacuum decay branching ratios (including electromagnetic ones) and elementary reactions, such as photo absorption on the nucleon and nuclei. The empirical information becomes scarce for highly excited resonances, but recent investigations have found their contributions to thermal dilepton and photon production rather moderate, at the level of tens of percent, leaving little room for major variations in the currently established emission rates of the hadronic matter.

Electromagnetic radiation may help to clarify the puzzling observations in small systems, i.e., whether the interpretations of hadronic observables (such as the sizeable elliptic flow coefficient together with rather modest modifications in the nuclear modification factor) require the presence of a QGP fireball, or are due to the prevalence of initial-state effects and/or finite-size escape mechanisms. The volume-type emission of *thermal* radiation from a would-be QGP fireball is expected to provide excess signals of several tens of percent in proton-nucleus collisions, which may well be measurable. An explicit implementation of photon emission processes into kinetic transport models could discriminate between the origin of elliptic flow in proton-nucleus collisions from a fully collective emission scenario of a QGP fireball.

The coherent emission of photons and dileptons due to the strong initial EM fields of the approaching nuclei (even without strong-interaction overlap) has recently become experimentally accessible, offering several opportunities. This radiation is characterized by a steep slope in the pair transverse momentum spectrum which is markedly softer than the typical scale, of the order of 200 MeV, of thermal radiation. Also, its centrality dependence is expected to be quite different. Recent STAR measurements indicate that a quantitative theoretical understanding of the observed excess at very low transverse still eludes us. In addition, ATLAS measurements of the acoplanarity of the back-to-back dimuons suggest a significant deflection mechanism. Quantitative theoretical analyses are required to determine to what extent or in fact even if[1] these new measurements may be used to diagnose the magnitude and duration of initial magnetic fields, and the role of multiple scattering of the leptons traversing the fireball.

## Soft Precision Physics

The study of particle production at low transverse momenta and its hadrochemical composition has repeatedly provided qualitatively novel insights and surprises. For instance, recent studies of hadrochemical composition as a function of event multiplicity across all system sizes from proton-proton to proton-nucleus to nucleus-nucleon show that high-multiplicity proton-proton collisions display heavy-ion like behavior in their strangeness composition. This indicates that there are more physics mechanisms operational in proton-proton collisions than traditionally thought. For instance, recent detailed analyses of charmonium production at low transverse momenta give support to a recombination mechanism in hadronization that deservers further scrutiny. Based on recent experience, one may expect that also in the future the study of soft particle production, if pursued with increased precision and over wider systematics, can contribute important insights. Our discussion of these topics focused on two so-far unexplored opportunities:

1. Fluctuations of conserved charges

---

[1] a recent preprint explains the impact parameter dependent broadening of the lepton pair transverse momentum in QED pair creation.

Fluctuations of conserved charges, in particular the net-baryon density, allow for the experimental identification of the existence of the (pseudo-critical) cross-over transition as predicted by lattice QCD. Their measurements are thus of fundamental interest. However experimental feasibility as well as various conceptual questions need to be addressed to arrive at measurements that can be unambiguously related to properties of the QCD phase transition. In particular:

- A prerequisite for establishing experimentally the existence of the pseudo-critical transition is the challenging measurement of cumulants of at least sixth order but likely eighth order. The statistics required for such a study (which will be substantial) needs to be estimated. Such an estimate would benefit from constraints on the probability distribution based on information from lattice QCD. At the very least it would be tremendously valuable to know if the signal sits in the tails or more in the center of the distribution.
- A faithful comparison of experimental measurements with lattice QCD requires an extrapolation to full phase space. In this limit, the fluctuation of conserved charges vanishes. Therefore, a reliable correction for charge conservation needs to be applied to data collected in a finite section of phase-space. In addition, in order to assure that the available acceptance in the present detector systems are sufficient a reliable estimate of the correlation length in momentum space is needed.
- A measurement of a baseline in proton-proton and proton-nucleus collisions will be essential for the interpretation of the nucleus-nucleus results.
- Most observables are consistent with viscous hydrodynamic evolution. Therefore, contrary to lattice simulations, the system created in a heavy-ion collision is not in static, global equilibrium. Instead we have a system that expands and at best is close to local equilibrium. To date it is not known to which extent the dynamical evolution affects the measurement of the various cumulants of conserved charges and this issue needs to be addressed with high priority.
- With increasing order of the cumulants one becomes more sensitive to small backgrounds. For example, although the net baryon density of the system created at the LHC is consistent with zero, the probability that some of the target and projectile nucleons get stopped is likely non-vanishing and may contaminate the signal. The same is true for the baryon from jets which may contribute to the net-baryon number at low transverse moments. Both these and other effects need to be studied.

## 2. (Anti-)(hyper-)nuclei and exotica

The measurements of complex anti- and hyper-nuclei in proton-proton, proton-nucleus, and nucleus-nucleus collisions do not only provide insights into the properties of the medium formed in these interactions, but are interesting for the following reasons:

- Binding energies and lifetimes of (anti-)hyper-nuclei are only poorly known and should be measured with high priority with the available and expected future statistics at RHIC and LHC. These measurements are also closely connected to the poorly constrained wave-functions of these objects which are discussed in the following.
- Measurements of the production yields as a function of the system size (charged particle multiplicity) in proton-proton, proton-nucleus, and nucleus-nucleus collisions will help clarify their currently not fully understood production mechanism. If these objects are produced by a coalescence processes, their production yields are sensitive to the extension of the wave-function with respect to the size of the emitting medium. Such a dependence is not expected in thermal-statistical models which give a very good description of the yields in central heavy-ion collisions. In principle, also in thermal-statistical calculations a corresponding form factor would enter which is currently not considered in the available models. Its influence is most likely

- negligible for light (anti-)nuclei, but it could turn out to be relevant for more extended objects such as hyper-nuclei.
- These measurements can serve as a testing ground for future measurements aimed at the understanding of exotic QCD objects whose internal structure (composite object vs. a true multi-quark states) is not known yet and might be clarified in the future with analogous measurements. Heavy-ion collisions might offer a unique possibility to constrain the wave functions of these objects.
- Beyond the fields of heavy-ion and accelerator-based physics, these measurements also provide crucial input for current (for instance AMS-02) and future (for instance AMS-100) searches for anti-matter with space-borne experiments. The accelerator-based experimental program should not be driven by current needs and understanding, but should rather be a systematic exploitation of all observable states at the currently available experimental facilities which provide a unique possibility that might not be available again in future.

Both for the distinction between coalescence and thermal-statistical production as well as for the studies related to astrophysics, a systematic and complete set of measurements covering a large set of objects (with different masses, spin states, binding energies, and wave functions) is thus desirable. In addition to transverse momentum- and ideally also rapidity-differential spectra, measurements of the azimuthal anisotropies in the production of these objects (elliptic flow) are needed to clarify the production mechanisms.

[1] "*Thoughts on heavy-ion physics in the high luminosity era: the soft sector*", 2nd Mt. St. Odile Meeting, F. Antinori et al., arXiv:1604.03310.